\documentclass [a4paper,fleqn, 12pt]{article}
\usepackage{graphicx}
\usepackage[small]{subfigure,epsfig}

\usepackage {amsmath} \usepackage{amssymb} \usepackage{cite}

\begin{document}

\title
{Logistic function as solution of many nonlinear differential equations}

\author{Nikolai A. Kudryashov, \and Mikhail A. Chmykhov}

\date{Department of Applied Mathematics, \\
National Research Nuclear University MEPhI \\
(Moscow Engineering Physics Institute), \\
31 Kashirskoe Shosse, 115409 Moscow, Russian Federation}


\maketitle

\begin{abstract}
The logistic function is shown to be solution of the Riccati equation, some second-order nonlinear ordinary differential equations and many third-order nonlinear ordinary differential equations. The list of the differential equations having solution in the form of the logistic function is presented.  The simple method of finding exact solutions of nonlinear partial differential equations (PDEs) is introduced. The essence of the method is based on  comparison of nonlinear differential equations obtained from PDEs with standard differential equations having solution in the form of the logistic function. The wide application of the logistic function for finding exact solutions of nonlinear differential equations is demonstrated.

\end{abstract}



\section{Introduction}

The logistic function (the sigmoid function) is determined by the following formula \cite{Gershenfield, Richards, KudrAMC13}
\begin{equation}
Q(z)=\frac{1}{1+e^{-z}},
\label{I_1}
\end{equation}
where $z$ is independent variable on the complex plane. We see that the logistic function has the pole of the first order on complex plane. This function can be used for finding exact solutions of nonlinear differential equations \cite{KudrAMC13, Kudr12a, KudrAMC13B}. Other variants of this approach without the logistic function were used in some papers before (see, for a example\cite{Kudr88, Kudr90, Kudr05a, Kudr08a, Kudr10a}).

One can see that the logistic function is the solution of the first order differential equation called the Riccati equation \cite{KudrAMC13, Kudr12a, KudrAMC13B}
\begin{equation}\begin{gathered}
Q_z-Q+Q^2=0.
\label{I_2}
\end{gathered}\end{equation}

The logistic function \eqref{I_1} can be presented taking the hyperbolic tangent into account because of the following formula
\begin{equation}\begin{gathered}
\frac{1}{1+e^{-z}}=\frac12\,\tanh{\left(\frac{z}{2}\right)}+\frac12.
\label{I_3}
\end{gathered}\end{equation}

However the logistic function is more convenient for finding exact solutions as it has been   illustrated in recent papers \cite{KudrAMC13, Kudr12a, KudrAMC13B}.

Let us show that the general solution of the Riccati equation can be expressed via the logistic function. The Riccati equation takes the form
\begin{equation}\begin{gathered}
y_z=a\,y^2+b\,y+c,
\label{I_4}
\end{gathered}\end{equation}
where $a$, $b$ and $c$ are arbitrary constants.

It is easy to obtain that the general solution of equation \eqref{I_4} can be written by means of formula
\begin{equation}\begin{gathered}
y=B-\frac{2\,B+b}{a}\,Q(z)\,\quad z=\frac{z^{'}-z_0}{2\,B+b},
\label{I_5}
\end{gathered}\end{equation}
where $z_0$ is an arbitrary constant,  $B$ is defined via constants $a$, $b$ and $c$ from the algebraic equation
\begin{equation}\begin{gathered}
a\,B^2+b\,B+c=0.
\label{I_6}
\end{gathered}\end{equation}
So, the logistic function is the solution of the Riccati equation to within transformations \eqref{I_5}.

The aim of this paper is to find some nonlinear ordinary differential equations of the second and the third order with exact solutions in the form of the logistic function and to show that there are  nonlinear partial differential equations having solution in the form of the logistic function.
We also illustrate that one can find exact solutions of many nonlinear partial differential equations using the list of standard nonlinear ordinary differential equations.

\section{Nonlinear ordinary differential equation of the second order with solution in form of logistic function}

Differentiating equation \eqref{I_2} with respect to $z$ we have the following second-order differential equation
\begin{equation}\begin{gathered}
Q_{zz}-Q_z+2\,Q\,Q_z=0.
\label{N_1}
\end{gathered}\end{equation}

It is obvious that the logistic function $Q(z)$ satisfies equation \eqref{N_1} as well. At that time if we use the equality
\begin{equation}\begin{gathered}
Q_z=Q-Q^2,
\label{N_2}
\end{gathered}\end{equation}
we obtain three other differential equation
\begin{equation}\begin{gathered}
Q_{zz}-Q+Q^2+2\,Q\,Q_z=0,
\label{N_3}
\end{gathered}\end{equation}
\begin{equation}\begin{gathered}
Q_{zz}-Q_z+2\,Q^2-2\,Q^3=0,
\label{N_4}
\end{gathered}\end{equation}
\begin{equation}\begin{gathered}
Q_{zz}-Q+3\,Q^2-2\,Q^3=0
\label{N_5}
\end{gathered}\end{equation}
having solutions in the form of the logistic function.

Taking into account these equations we can present the other second order nonlinear ordinary differential equations with solutions in the form of logistic function. These equations take the form
\begin{equation}\begin{gathered}
Q_{zz}-Q_z+2\,Q\,Q_z+F_1(Q, \,Q_z, \ldots)\,(Q_z-Q+Q^2)=0,
\label{N_6}
\end{gathered}\end{equation}
\begin{equation}\begin{gathered}
Q_{zz}-Q+Q^2+2\,Q\,Q_z+F_2(Q,\,Q_z, \ldots)\,(Q_z-Q+Q^2)=0,
\label{N_7}
\end{gathered}\end{equation}
\begin{equation}\begin{gathered}
Q_{zz}-Q_z+2\,Q^2-2\,Q^3+F_3(Q,\,Q_z, \ldots)\,(Q_z-Q+Q^2)=0,
\label{N_8}
\end{gathered}\end{equation}
\begin{equation}\begin{gathered}
Q_{zz}-Q+3\,Q^2-2\,Q^3+F_4(Q,\,Q_z, \ldots)\,(Q_z-Q+Q^2)=0,
\label{N_9}
\end{gathered}\end{equation}
where $F_j(Q,\,Q_z, \ldots),\quad j=1,\ldots,4$ are some dependencies on $Q$, $Q_z$ and so on.

Let us call equations \eqref{N_6} - \eqref{N_9} as the standard nonlinear differential equations.

In the present time there are a lot of different methods for finding exact solutions of nonlinear differential equations. Here we only call some of them: the singular manifold method \cite{Kudr88, Kudr90} that was modified by Kudryashov \cite{Kudr12a, KudrAMC13, Kabir01,Kabir02}, the simplest equation method and some its modifications \cite{Kudr05a, Kudr08a, Kudr10a, Biswas, Vitanov01, Vitanov02, Vitanov03}, the tanh-method \cite{Malfliet, Hereman01, Parkes01, Hereman02, Hereman03, Biswas01}, the $G^{'}/G$ - expansion method \cite{Wang, Kudr09k, Kudr10k}, the method for finding meromorphic solutins of nonlinear differential equations \cite{Kudr10n, Kudr10ss, Kudr11b}.

However we can look for exact solutions of many nonlinear partial differential equations taking into account the list of standard equations \eqref{N_6} - \eqref{N_9} using the following simple algorithm. Let us take the nonlinear partial differential equation with the solution of the first order pole
\begin{equation}\begin{gathered}
E_1(u,u_t,u_x,u_{xx}, \ldots)=0
\label{N_{10}}
\end{gathered}\end{equation}
Using the traveling wave solution
\begin{equation}\begin{gathered}
u(x,t)=y(z),\quad z=k\,x-\omega\,t-\,k\,x_0
\label{N_{11}}
\end{gathered}\end{equation}
we have the nonlinear ordinary differential equation in the form
\begin{equation}\begin{gathered}
E_2(y,-\omega\,y_z,k\,y_z,\,k^2\,y_{zz}, \ldots)=0
\label{N_{10}}
\end{gathered}\end{equation}

Using $y=a_0\,Q(z)$ in \eqref{N_{10}} and comparing new form of equation \eqref{N_{10}} with the standard equation of the list   \eqref{N_6} - \eqref{N_9} we can find a solution in the form of the logistic function. Let us demonstrate this algorithm taking into account some examples.

\section{The logistic function as a solution of the Burgers equation}

Let us illustrate that the logistic function is a solution of the Burgers equation. The Burgers equation can be written in the form
\begin{equation}
u_t+2\,u\,u_x=\mu\,u_{xx}.
\label{B_1}
\end{equation}

Assuming
\begin{equation}
u(x,t)=y(z), \qquad z=k\,x-\omega\,t-k\,x_0
\label{B_2}
\end{equation}
we have the nonlinear ordinary differential equation in the form
\begin{equation}
k^2\,y_{zz}+\omega\,y_z-2\,k\,y\,y_z=0.
\label{B_3}
\end{equation}

From the list of standard equations we take equation \eqref{N_6} at $F_1(Q)=0$
\begin{equation}
Q_{zz}-Q_z+2\,Q\,Q_z=0.
\label{B_4}
\end{equation}

One can note that equations \eqref{B_3} and \eqref{B_4}  are the same  in the case $\omega=-k^2$ and $y=-k\,Q(z)$ in \eqref{B_3}.  As a result of the comparison we find the solution of the Burgers equation \eqref{B_1} in the form
\begin{equation}
u(x,t)=-\frac{k}{1+\exp{(k\,x-k^2\,t-k\,x_0)}}.
\label{B_4}
\end{equation}

We obtain that the  solution of the Burgers equation is expressed via the logistic function.

\section{The logistic function as a solution of the Huxley equation}

The Huxley equation takes the form
\begin{equation}
u_t=\,u_{xx}+u\,(\alpha-u)\,(u-1).
\label{Hu_1}
\end{equation}

Assuming
\begin{equation}
u(x,t)=y(z), \qquad z=k\,x-\omega\,t-k\,x_0
\label{Hu_2}
\end{equation}
we obtain the nonlinear ordinary differential equation
\begin{equation}
k^2\,y_{zz}+\omega\,y_z-\alpha\,y+(\alpha+1)\,y^2-y^3=0.
\label{Hu_3}
\end{equation}

From \eqref{N_9} we have the standard equation at $F_4=m$ in the form
\begin{equation}
Q_{zz}+m\,Q_z-(m+1)\,Q+(m+3)\,Q^2-2\,Q^3=0,
\label{Hu_4}
\end{equation}
where $m$ is the unknown parameter.

Assuming $y(z)=a_0\,Q(z)$ and comparing coefficients of equations \eqref{Hu_3} and \eqref{Hu_4} we obtain the following algebraic equations
\begin{equation}
\omega=m\,k^2,\quad \alpha=(m+1)\,k^2, \quad a_0\,(\alpha+1)=(m+3)\,k^2,\quad a_0^2=2\,k^2
\label{Hu_5}
\end{equation}

Solving equations \eqref{Hu_5} we find values of parameters
\begin{equation}
a_0=k\,\sqrt{2},\quad k=\frac{\alpha}{\sqrt{2}}\, \quad \omega=k\,\sqrt{2}-k^2,\quad m=\frac{\sqrt{2}}{k}-1.
\label{Hu_6}
\end{equation}

So, there is a solution of the Huxley equation \eqref{Hu_1} expressed via the logistic function in the form
\begin{equation}
u(x,t)=\,\frac{\alpha}{1+\exp{\left(-\frac{\alpha}{\sqrt{2}}\,x+
\frac{\alpha}{\sqrt{2}}\,x_0-\frac{\alpha^2}{4}\,t+\,\alpha\,t\right)}}
\label{Hu_7}
\end{equation}
where $x_0$ is an arbitrary constant.

\section{The logistic function as a solution of the Burgers -- Fisher equation}

Let us consider the Burgers-Fisher equation in the form
\begin{equation}
u_t+\,u\,u_x=\,u_{xx}-\beta\,u-\delta\,u^2.
\label{H_1}
\end{equation}

Assuming
\begin{equation}
u(x,t)=y(z), \qquad z=k\,x-\omega\,t
\label{H_2}
\end{equation}
we obtain the nonlinear ordinary differential equation
\begin{equation}
k^2\,y_{zz}-\,k\,y\,y_z+\omega\,y_z-\beta\,y-\delta\,y^2=0.
\label{H_3}
\end{equation}
Assuming $F_1(Q)=2$ in \eqref{N_6}
we have the standard equation in the form
\begin{equation}
Q_{zz}+2\,Q\,Q_z+\,Q_z-2\,Q+2\,Q^2=0.
\label{H_4}
\end{equation}

Taking $y=a_0\,Q(z)$ in \eqref{H_3} and comparing two equations \eqref{H_3} and \eqref{H_4} we have the following algebraic equations
\begin{equation}\begin{gathered}
a_0\,=-2\,k, \quad \omega=k^2,\quad \beta=2\,k^2,\quad a_0\,\delta=-2\,k^2.
\label{H_5}
\end{gathered}\end{equation}

Solving last equations we obtain the following values parameters
\begin{equation}\begin{gathered}
k=\sqrt{\frac{\beta}{2}},\quad \omega=\frac{\beta}{2},\quad a_0=-\sqrt{2\,\beta},\quad \delta=\sqrt{\frac{\beta}{2}}.
\label{H_6}
\end{gathered}\end{equation}

The solution of equation \eqref{H_1} in the form of the logistic function can be written in the form
\begin{equation}\begin{gathered}
u(x,t)=-
\frac{\sqrt{2\,\beta}}{1+\exp{\left(-\sqrt{\frac{\beta}{2}}\,x+\frac{\beta}{2}\,t+
\sqrt{\frac{\beta}{2}}\,x_0\right)}},
\label{H_8}
\end{gathered}\end{equation}
where $x_0$ is an arbitrary constant.

\section{The logistic function as a solution of the modified Korteweg-de Vries equation with dissipation}

The modified Korteweg-de Vries equation with dissipation can be written as the following
\begin{equation}
u_t+\beta\,u^2\,u_x+u_{xxx}=\mu\,u_{xx}.
\label{G_1}
\end{equation}

Taking the traveling wave solutions
\begin{equation}
u(x,t)=y(z), \qquad z=k\,x-\omega\,t
\label{G_2}
\end{equation}
in \eqref{G_1} we get the nonlinear ordinary differential equation in the form
\begin{equation}
k^3\,y_{zzz}-\mu\,k^2\,u_{xx}-\omega\,y_z+\beta\,k\,y^2\,y_z=0.
\label{G_3}
\end{equation}

After integration of \eqref{G_3} with respect to $z$ we have the second-order nonlinear ordinary differential equation
\begin{equation}
k^3y_{zz}-\mu\,k^2\,y_z\,-\omega\,y+\frac13\,\beta\,k\,y^3=C_1,
\label{G_4}
\end{equation}
where $C_1$ is a constant of integration.

Take the standard equation in the form \eqref{N_4} at $F_3(Q)=0$
\begin{equation}
Q_{zz}-3\,Q_z+2\,Q-2\,Q^3=0
\label{G_5}
\end{equation}

Substituting $y=a_0\,Q(z)$ and comparing equations \eqref{G_4} and \eqref{G_5} we obtain
\begin{equation}
C_1=0, \quad k=\frac{\mu}{3}\,\quad a_0=\frac{\mu}{3},\quad \omega=-2\,k^3,\quad \beta=-6.
\label{G_6}
\end{equation}
Solution of the modified Korteweg-de Vries equation \eqref{G_1}  with dissipation can be written in the form
\begin{equation}
u(x,t)=\frac{\mu}{3+3\,\exp{\left(-\frac{\mu}{3}\,x-\frac{2\,\mu^3}{27}\,t+\frac{\mu}{3}\,x_0\right)}},
\label{G_7}
\end{equation}
where $x_0$ is arbitrary constant.

\section{The third order nonlinear differential equation having solution in the form of logistic function}

Differentiating \eqref{N_1} and \eqref{N_3} - \eqref{N_5} with respect to $z$ and taking into account the Riccati equation in the form \eqref{I_3}, we find 60 nonlinear ordinary differential equations of the third order with solutions in the form of the logistic function. These 60 equations can be written as  the following list
\begin{equation}
Q_{zzz}=Q_{zz}-2\,{Q_{z}}^2-2\,Q\,Q_{{zz}},
\label{6.1}
\end{equation}
\begin{equation}
Q_{zzz}=Q_{zz}-2\,{Q_{{z}}}^{2}-2\,Q\,Q_{{z}}+4\,{Q}^{3}-4\,{Q}^{4},
 \label{6.2}
\end{equation}
\begin{equation}
Q_{zzz}=Q_{{{zz}}}-2\,{Q_{{z}}}^{2}-2\,Q\,Q_{{z}}+4\,{Q}^{2}\,Q_{{z}},
\label{6.3}
\end{equation}
\begin{equation}
Q_{zzz}=Q_{{{zz}}}-2\,{Q_{{z}}}^{2}-2\,{Q}^{2}+2\,{Q}^{3}+4\,{Q}^{2}\,Q_{{z}}
\label{6.4}
\end{equation}
\begin{equation}
Q_{zzz}=Q_{{{zz}}}-2\,{Q_{{z}}}^{2}-2\,{Q}^{2}+6\,{Q}^{3}-4\,{Q}^{4}
\label{6.5}
\end{equation}
\begin{equation}
Q_{zzz}=Q_{{zz}}-2\,Q^2+4\,{Q}^{3}-2\,Q^4-2\,Q\,Q_{{zz}},
\label{6.6}
\end{equation}
\begin{equation}
Q_{zzz}=Q_{{{zz}}}-2\,{Q}^{2}+8\,{Q}^{3}-6\,{Q}^{4}-2\,QQ_{{z}},
\label{6.7}
\end{equation}
\begin{equation}
Q_{zzz}=Q_{{{zz}}}-2\,{Q}^{2}+4\,{Q}^{3}-2\,{Q}^{4}-2\,QQ_{{z}}+4\,{Q}^{2}\,
Q_{{z}},
 \label{6.8}
\end{equation}
\begin{equation}
Q_{zzz}=Q_{{{zz}}}-4\,{Q}^{2}+4\,{Q}^{3}+6\,{Q}^{2}\,Q_{{z}},
\label{6.9}
\end{equation}
\begin{equation}
Q_{zzz}=Q_{{{zz}}}-4\,{Q}^{2}+10\,{Q}^{3}-6\,{Q}^{4},
\label{6.10}
\end{equation}
\begin{equation}
Q_{zzz}=Q_{{{zz}}}-4\,{Q}^{2}+6\,{Q}^{3}-2\,{Q}^{4}+4\,{Q}^{2}Q_{{z}},
 \label{6.11}
\end{equation}
\begin{equation}
Q_{zzz}=Q_{{zz}}-2\,Q\,Q_{{z}}+2\,{Q}^{2}\,Q_{{z}}-2\,Q\,Q_{{{zz}}},
\label{6.12}
\end{equation}
\begin{equation}
Q_{zzz}=Q_{{zz}}-2\,Q^2+2\,Q^3+2\,{Q}^{2}\,Q_{{z}}-2\,Q\,Q_{{{zz}}},
\label{6.13}
\end{equation}
\begin{equation}
Q_{zzz}=Q_{{{zz}}}-4\,Q\,Q_{{z}}+6\,{Q}^{3}-6\,{Q}^{4},
\label{6.14}
\end{equation}
\begin{equation}
Q_{zzz}=Q_{{{zz}}}+2\,{Q}^{2}\,Q_{{z}}-4\,Q\,Q_{{z}}+4\,{Q}^{3}-4\,{Q}^{4}
\label{6.15}
\end{equation}
\begin{equation}
Q_{zzz}=Q_{{{zz}}}+6\,{Q}^{2}\,Q_{{z}}-4\,Q\,Q_{{z}},
 \label{6.16}
\end{equation}
\begin{equation}
Q_{zzz}=Q_{z}-2\,{Q}^{2}+2\,{Q}^{3}-2\,Q_{z}^{2}-2\,Q\,Q_{zz},
\label{6.17}
\end{equation}
\begin{equation}
Q_{zzz}=Q_{{z}}-2\,{Q}^{2}+2\,{Q}^{3}-2\,Q\,Q_{{z}}+2\,{Q}^{2}Q_{{z}}-2\,Q\,Q_{{{zz}}},
\label{6.18}
\end{equation}
\begin{equation}
Q_{zzz}=Q_{{z}}-2\,{Q}^{2}+4\,{Q}^{3}-2\,{Q}^{4}-2\,Q\,Q_{{z}}-2\,Q\,Q_{{{zz}}},
\label{6.19}
\end{equation}
\begin{equation}
Q_{zzz}=Q_{{z}}-4\,{Q}^{2}+6\,{Q}^{3}-2\,{Q}^{4}-2\,Q\,Q_{{{zz}}},
\label{6.20}
\end{equation}
\begin{equation}
Q_{zzz}=Q_{{z}}-6\,{Q}^{2}+12\,{Q}^{3}-6\,{Q}^{4}
 \label{6.21}
\end{equation}
\begin{equation}
Q_{zzz}=Q_{{z}}-2\,Q\,Q_{{z}}-2\,{Q_{{z}}}^{2}-2\,Q\,Q_{{zz}},
\label{6.22}
\end{equation}
\begin{equation}
Q_{zzz}=Q_{{z}}-2\,Q\,Q_{{z}}-2\,{Q_{{z}}}^{2}-2\,{Q}^{2}+2\,{Q}^{3}+4\,{Q}^{2}Q
_{{z}}
\label{6.23}
\end{equation}
\begin{equation}
Q_{zzz}=Q_{{z}}-2\,Q\,Q_{{z}}+4\,{Q}^{2}Q_{{z}}-4\,{Q}^{2}+6\,{Q}^{3}-2\,{Q}^{4}
 \label{6.24}
\end{equation}
\begin{equation}
Q_{zzz}=Q_{{z}}-2\,Q\,Q_{{z}}-2\,{Q_{{z}}}^{2}-2\,{Q}^{2}+6\,{Q}^{3}-4\,{Q}^{4}
\label{6.25}
\end{equation}
\begin{equation}
Q_{zzz}=Q_{{z}}-2\,Q\,Q_{{z}}-4\,{Q}^{2}+10\,{Q}^{3}-6\,{Q}^{4}
\label{6.26}
\end{equation}
\begin{equation}
Q_{zzz}=Q_{{z}}-4\,Q\,Q_{{z}}+2\,{Q}^{2}\,Q_{{z}}-2\,Q\,Q_{{{zz}}},
\label{6.27}
\end{equation}
\begin{equation}
Q_{zzz}=Q_{{z}}-4\,Q\,Q_{{z}}-2\,{Q_{{z}}}^{2}+4\,{Q}^{2}Q_{{z}}
 \label{6.28}
\end{equation}
\begin{equation}
Q_{zzz}=Q_{{z}}-4\,Q\,Q_{{z}}+4\,{Q}^{2}\,Q_{{z}}-2\,{Q}^{2}+4\,{Q}^{3}-2\,{Q}^{4},
\label{6.29}
\end{equation}
\begin{equation}
Q_{zzz}=Q_{{z}}-4\,Q\,Q_{{z}}-2\,{Q_{{z}}}^{2}+4\,{Q}^{3}-4\,{Q}^{4},
 \label{6.30}
\end{equation}
\begin{equation}
Q_{zzz}=Q_{{z}}-4\,Q\,Q_{{z}}-2\,{Q}^{2}+8\,{Q}^{3}-6\,{Q}^{4}
 \label{6.31}
\end{equation}
\begin{equation}
Q_{zzz}=Q_{{z}}-4\,QQ_{{z}}+6\,{Q}^{2}Q_{{z}}-2\,{Q}^{2}+2\,{Q}^{3}
\label{6.32}
\end{equation}
\begin{equation}
Q_{zzz}=Q_{{z}}-6\,Q\,Q_{{z}}+6\,{Q}^{2}Q_{{z}},
 \label{6.33}
\end{equation}
\begin{equation}
Q_{zzz}=Q_{{z}}-6\,Q\,Q_{{z}}+6\,{Q}^{3}-6\,{Q}^{4},
\label{6.34}
\end{equation}
\begin{equation}
Q_{zzz}=Q_{{z}}+4\,{Q}^{2}\,Q_{{z}}-4\,{Q}^{2}+4\,{Q}^{3}-2\,{Q_{{z}}}^{2},
 \label{6.35}
\end{equation}
\begin{equation}
Q_{zzz}=Q_{{z}}+4\,{Q}^{2}Q_{{z}}-6\,{Q}^{2}+8\,{Q}^{3}-2\,{Q}^{4}
\label{6.36}
\end{equation}
\begin{equation}
Q_{zzz}=Q_{{z}}+6\,{Q}^{2}\,Q_{{z}}-6\,{Q}^{2}+6\,{Q}^{3},
 \label{6.37}
\end{equation}
\begin{equation}
Q_{zzz}=Q_z-6\,{Q_{{z}}}^{2},
\label{6.38}
\end{equation}
\begin{equation}
Q_{zzz}=Q_{{z}}-2\,{Q_{{z}}}^{2}-4\,{Q}^{2}+8\,{Q}^{3}-4\,{Q}^{4}
 \label{6.39}
\end{equation}
\begin{equation}
\label{6,40}
Q_{zzz}=Q-{Q}^{2}-2\,Q\,Q_{{z}}-2\,{Q_{{z}}}^{2}-2\,QQ_{{{zz}}},
\end{equation}
\begin{equation}
Q_{zzz}=Q-3\,{Q}^{2}+2\,{Q}^{3}-2\,{Q_{{z}}}^{2}-2\,Q\,Q_{{{zz}}},
\label{6.41}
\end{equation}
\begin{equation}
Q_{zzz}=Q-3\,{Q}^{2}+4\,{Q}^{3}-2\,{Q}^{4}-2\,Q\,Q_{{z}}-2\,Q\,Q_{{{zz}}},
\label{6.42}
\end{equation}
\begin{equation}
Q_{zzz}=Q-5\,{Q}^{2}+6\,{Q}^{3}-2\,{Q}^{4}-2\,Q\,Q_{{{zz}}},
\label{6.43}
\end{equation}
\begin{equation}
Q_{zzz}=Q-{Q}^{2}-6\,{Q_{{z}}}^{2},
\label{6.44}
\end{equation}
\begin{equation}
Q_{zzz}=Q-{Q}^{2}+6\,{Q}^{3}-6\,{Q}^{4}-6\,Q\,Q_{{z}},
\label{6.45}
\end{equation}
\begin{equation}
Q_{zzz}=Q-{Q}^{2}-4\,QQ_{{z}}-2\,{Q_{{z}}}^{2}+4\,{Q}^{2}Q_{{z}},
\label{6.46}
\end{equation}
\begin{equation}
Q_{zzz}=Q-{Q}^{2}+4\,{Q}^{3}-4\,{Q}^{4}-2\,{Q_{{z}}}^{2}-4\,Q\,Q_{{z}}
 \label{6.47}
\end{equation}
\begin{equation}
Q_{zzz}=Q-{Q}^{2}-6\,Q\,Q_{{z}}+6\,{Q}^{2}Q_{{z}}
 \label{6.48}
\end{equation}
\begin{equation}
Q_{zzz}=Q-3\,{Q}^{2}+8\,{Q}^{3}-6\,{Q}^{4}-4\,Q\,Q_{{z}}
\label{6.49}
\end{equation}
\begin{equation}
Q_{zzz}=Q-3\,{Q}^{2}+4\,{Q}^{3}-2\,{Q}^{4}+4\,{Q}^{2}\,Q_{{z}}-4\,Q\,Q_{{z}},
\label{6.50}
\end{equation}
\begin{equation}
Q_{zzz}=Q-3\,{Q}^{2}-2\,Q\,Q_{{z}}-2\,{Q_{{z}}}^{2}+2\,{Q}^{3}+4\,{Q}^{2}Q_{{z}}
\label{6.51}
\end{equation}
\begin{equation}
Q_{zzz}=Q-3\,{Q}^{2}-2\,Q\,Q_{{z}}-2\,{Q_{{z}}}^{2}+6\,{Q}^{3}-4\,{Q}^{4}
 \label{6.52}
\end{equation}
\begin{equation}
Q_{zzz}=Q-3\,{Q}^{2}+2\,{Q}^{3}-4\,Q\,Q_{{z}}+6\,{Q}^{2}Q_{{z}}
 \label{6.53}
\end{equation}
\begin{equation}
Q_{zzz}=Q-5\,{Q}^{2}+4\,{Q}^{3}-2\,{Q_{{z}}}^{2}+4\,{Q}^{2}\,Q_{{z}},
 \label{6.54}
\end{equation}
\begin{equation}
Q_{zzz}=Q-5\,{Q}^{2}+8\,{Q}^{3}-2\,{Q_{{z}}}^{2}-4\,{Q}^{4}
\label{6.55}
\end{equation}
\begin{equation}
Q_{zzz}=Q-5\,{Q}^{2}-2\,Q\,Q_{{z}}+6\,{Q}^{3}-2\,{Q}^{4}+4\,{Q}^{2}\,Q_{{z}}
 \label{6.56}
\end{equation}
\begin{equation}
Q_{zzz}=Q-5\,{Q}^{2}-2\,Q\,Q_{{z}}+10\,{Q}^{3}-6\,{Q}^{4}
 \label{6.57}
\end{equation}
\begin{equation}
Q_{zzz}=Q-7\,{Q}^{2}+12\,{Q}^{3}-6\,{Q}^{4}
 \label{6.58}
\end{equation}
\begin{equation}
Q_{zzz}=Q-7\,{Q}^{2}+6\,{Q}^{3}+6\,{Q}^{2}\,Q_{{z}},
 \label{6.59}
\end{equation}
\begin{equation}
Q_{zzz}=Q-7\,{Q}^{2}+8\,{Q}^{3}-2\,{Q}^{4}+4\,{Q}^{2}Q_{{z}}
 \label{6.60}
\end{equation}

We can suggest also other nonlinear ordinary differential equations of the third order taking into account equations \eqref{I_2} and the second-order nonlinear differential equations from section 2 because we always add or subtract from equations \eqref{6.1} -- \eqref{6.60} the following expressions
\begin{equation}
P_1(Q,Q_{z},\ldots)\,\left(Q_{zz}-Q_z-2Q\,Q_z\right)+F_1(Q,Q_{z},\ldots)\left(Q_z-Q+Q^2\right),
 \label{6.61}
\end{equation}
\begin{equation}
P_2(Q,Q_{z},\ldots)\,\left(Q_{zz}-Q+Q^2-2Q\,Q_z\right)+F_2(Q,Q_{z},\ldots)\left(Q_z-Q+Q^2\right),
 \label{6.62}
\end{equation}
\begin{equation}
P_3(Q,Q_{z},\ldots)\,\left(Q_{zz}-Q_z+2Q^2-2\,Q^3\right)+F_3(Q,Q_{z},\ldots)\left(Q_z-Q+Q^2\right),
 \label{6.63}
\end{equation}
\begin{equation}
P_4(Q,Q_{z},\ldots)\,\left(Q_{zz}+3\,Q^2-2\,Q^3\right)+F_4(Q,Q_{z},\ldots)\left(Q_z-Q+Q^2\right),
 \label{6.64}
\end{equation}
where $P_j(Q,\,Q_z,\,\ldots)$ $j=1,\ldots,4$ are some dependencies on $Q,\,Q_z,\,\ldots$.
Nonlinear differential equations with solutions in the from of the logistic function  can be useful for constructing exact solutions of some nonlinear partial differential equations.

\section{Logistic function as a solution of the Gardner equation}

The Gardner equations is the generalization of the modified Korteweg-de Vries equation and  takes the form
\begin{equation}
u_t+\alpha\,u\,u_x+\beta\,u^2\,u_x+u_{xxx}=0
 \label{G.1}
\end{equation}

Taking into account the traveling wave solutions
\begin{equation}
u(x,t)=y(z),\quad z=k\,x-\omega\,t-k\,x_0
 \label{G.2}
\end{equation}
we have from \eqref{G.1} the following equation
\begin{equation}
k^3\,y_{zzz}-\omega\,y_z+\alpha\,k\,y\,y_z+\beta\,k\,y^2\,y_z=0.
 \label{G.3}
\end{equation}

Assuming
\begin{equation}
\omega=k^3,\quad y=a_0\,Q(z)
 \label{G.4}
\end{equation}
we obtain equation \eqref{6.34} from the list of standard equations at
\begin{equation}
a_0=\frac{6\,k^2}{\alpha},\quad \beta=-\frac{\alpha^2}{6\,k^2}.
 \label{G.5}
\end{equation}
It takes the form
\begin{equation}
Q_{zzz}-Q_{z}+6\,Q\,Q_z-6\,Q^2\,Q_z=0.
 \label{G.6}
\end{equation}

The solution of equation \eqref{G_1} is expressed by formula
\begin{equation}
u(x,t)=\frac{6\,k^2}{\alpha(1+\exp{(k\,x-k^3-k\,x_0)})},
 \label{G.7}
\end{equation}
where $x_0$ is an arbitrary constant.

\section{Logistic function as a solution of the Korteweg-de Vries equation with source}

Let us demonstrate that the logistic function is a solution of the Korteweg-de Vries equation with source in the form
\begin{equation}
u_t+6\,u\,u_x+u_{xxx}+\beta\,u^2+\gamma\,u^3+\delta\,u^4=0.
 \label{MM.1}
\end{equation}

Using the traveling wave solutions in \eqref{MM.1}
\begin{equation}
u(x,t)=y(z),\quad z=k\,x-\omega\,t-k\,x_0
 \label{MM.2}
\end{equation}
we obtain  the equation from \eqref{MM.1} in the form
\begin{equation}
k^3\,y_{zzz}-\omega\,y_z+6\,k\,y\,y_z+\beta\,y^2+\gamma\,y^3+\delta\,y^4=0.
 \label{MM.3}
\end{equation}
Let us take equation \eqref{6.26} from our list
\begin{equation}
Q_{zzz}-Q_z+2\,Q\,Q_z+4\,Q^2-10\,Q^3+6\,Q^4=0.
 \label{MM.4}
\end{equation}

Assuming $y=a_0\,Q(z)$ and comparing equations \eqref{MM.3} and \eqref{MM.7} we have the algebraic equations
\begin{equation}
6\,a_0=2\,k^2,\quad \omega=k^3,\quad \beta\,a_0=4\,k^3, \quad \gamma\,a_0^2=-10\,k^3, \quad \delta\,a_0^3=6\,k^3
 \label{MM.5}
\end{equation}
we obtain the following values of the parameters
\begin{equation}
k=\frac{\beta}{12},\quad a_0=\frac{\beta^2}{432},\quad \omega=\frac{\beta^3}{1728},\quad\gamma=-\frac{1080}{\beta},\quad \delta=\frac{278936}{\beta^3}.
\label{MM.6}
\end{equation}

We have the exact solution of equation \eqref{MM.1} in the form
\begin{equation}
u(x,t)=\frac{\beta^2}{432+432\,\exp{\left(-\frac{\beta\,x}{12}+\frac{\beta^3\,t}{1728}+
\frac{\beta\,x_0}{12}\right)}},
 \label{MM.7}
\end{equation}
where $x_0$ is an arbitrary constant, $\gamma$ and $\delta$ are determined by expressions \eqref{MM.6}.

\section{Logistic function as a solution of the modified Korteweg--de Vries equation with souce}

Let us consider the modified Korteweg-de Vries equation with source and show that the logistic function is solution of this equation as well. The modified Korteweg-de Vries equation with source takes the form
\begin{equation}
u_t-6\,u^2\,u_x+u_{xxx}+\alpha\,u^2+\beta\,u^3=0
 \label{MG.1}
\end{equation}

Using the traveling wave solutions
\begin{equation}
u(x,t)=y(z),\quad z=k\,x-\omega\,t-k\,x_0
 \label{MG.2}
\end{equation}
we obtain from \eqref{MG.1} the following equation
\begin{equation}
k^3\,y_{zzz}-6\,k\,y^2\,y_z-\omega\,y_z+\alpha\,y^2+\beta\,y^3=0.
 \label{MG.3}
\end{equation}
Let us take standard equation \eqref{6.37} from the list of section 6 in the form
\begin{equation}
Q_{zzz}-6\,Q^2\,Q_z-Q_z+6\,Q^2-6\,Q^3
 \label{MG.4}
\end{equation}

Assuming $y=a_0\,Q(z)$ and comparing two equations we have the following equations
\begin{equation}
\omega=k^3,\quad a_0^2=k^2,\quad  \alpha\,a_0=6\,k^3, \quad a_0^2\,\beta=-6\,k^3.
\label{MG.5}
\end{equation}

Solving equations \eqref{MG.5} we obtain
\begin{equation}
a_0=\pm \,k, \quad k=-\frac{\beta}{6}, \quad \omega=-\frac{\beta^3}{216},\quad \alpha=\mp\,\frac{\beta^2}{6}.
\label{MG.6}
\end{equation}

We obtain the following  solution of equation \eqref{MG.1} at $\alpha=\mp\,\frac{\beta^2}{6}$
\begin{equation}
u(x,t)=\pm \frac{\,\beta}{6+6\,\exp{(\frac{\beta\,x}{6}+\frac{\beta^3\,t}{216}-\frac{\beta\,x_0}{6})}},
\label{MG.7}
\end{equation}
where $x_0$ is an arbitrary constant.

\section{Logistic function as a solution of the generalized Kuramoto-Sivashinsky equation}

Let us consider the generalized Kuramoto-Sivashinsky equation in the form
\begin{equation}
u_t+u_{xxxx}-\mu\,u_{xx}-\alpha\,u^2\,u_x+\beta\,u^3\,u_x=0
 \label{KS.1}
\end{equation}

Using the traveling wave solution
\begin{equation}
u(x,t)=y(z),\quad z=k\,x-\omega\,t-k\,x_0
 \label{KS.2}
\end{equation}
we have from \eqref{KS.1} the nonlinear ordinary differential equation
\begin{equation}
k^4\,y_{zzzz}-\mu\,k^2\,y_{zz}-\alpha\,k\,u^2\,u_z+\beta\,k\,u^3\,u_z-\omega\,y_z=0.
 \label{KS.3}
\end{equation}

After integration \eqref{KS.3} with respect to $z$ we obtain
\begin{equation}
k^4\,y_{zzz}-\mu\,k^2\,y_{z}-\frac13\,\alpha\,k\,u^3+\frac14\,\beta\,k\,u^4-\omega\,y=C_1.
 \label{KS.4}
\end{equation}

Taking into account the standard equation in the form
\begin{equation}
Q_{zzz}-7\,Q_z+6\,Q-12\,Q^3+6\,Q^4=0.
 \label{KS.5}
\end{equation}
The last equation can be found if we take equation \eqref{6.21} from our list and subtract equation of the first order
\[6\,(Q_z-Q+Q^2).\]

Comparing equations \eqref{KS.4} and \eqref{KS.5} we have the algebraic equations
\begin{equation}
C_1=0,\quad \mu=7\,k^2,\quad a_0^2\,\alpha=36\,k^3,\quad a_0^3\,\beta=24\,k^3,\quad \omega=-6\,k^4
 \label{KS.6}
\end{equation}

Solving equations \eqref{KS.6} we obtain the following values of parameters
\begin{equation}
a_0=\frac{2\alpha}{3\,\beta},\quad k=\frac{\alpha}{(9\,\beta)^{2/3}},\quad  \mu=\,\frac{7\alpha^2}{(9\,\beta)^{4/3}},\quad \omega=-\frac{6\,\alpha^4}{{9\,\beta}^{8/3}}
 \label{KS.7}
\end{equation}

We have the following exact solution of equation \eqref{KS.1} at $\mu=\,\frac{7\alpha^2}{(9\,\beta)^{4/3}}$
\begin{equation}
u(x,t)=\frac{2\,\alpha}{3\,\beta+3\,\beta\,\exp{\left(-\frac{\alpha\,x-\alpha\,x_0}{(9\,\beta)^{2/3}}-
\frac{6\,\alpha^4\,t}{{9\,\beta}^{8/3}}\right)}},
 \label{KS.8}
\end{equation}
where $x_0$ is an arbitrary constant.

\section{Logistic function as a solution of the Korteweg-de Vries equation}

We demonstrate the effect of our approach for nonlinear partial differential equations with solutions having first order pole. However the method is working for other nonlinear differential equations too. Let us illustrate this one using the famous Korteweg-de Vries equation
\begin{equation}\begin{gathered}
u_t+6\,u\,u_x+u_{xxx}=0.
\label{KdV_1}
\end{gathered}\end{equation}

Using the traveling wave solutions
\begin{equation}\begin{gathered}
u(x,t)=y(z),\quad z=k\,x-\omega\,t
\label{KdV_2}
\end{gathered}\end{equation}
we have from equation \eqref{KdV_1}
\begin{equation}\begin{gathered}
k^3\,y_{zzz}+6\,k\,y\,y_z-\omega\,y_z=0.
\label{KdV_3}
\end{gathered}\end{equation}

After integration equation \eqref{KdV_3} with respect to $z$ we get
\begin{equation}\begin{gathered}
k^3\,y_{zz}+3\,k\,y^2-\omega\,y=C_1.
\label{KdV_4}
\end{gathered}\end{equation}

Let us take \eqref{6.38} as the standard equation
\begin{equation}\begin{gathered}
Q_{zzz}=Q_z-6\,Q_z^2.
\label{KdV_5}
\end{gathered}\end{equation}

Taking into account new variable $V=Q_z$ we obtain the standard equation for variable $V(z)$
\begin{equation}\begin{gathered}
V_{zz} + 6\,V^2-V=0.
\label{KdV_6}
\end{gathered}\end{equation}

Assuming $y=a_0\,V(z)$ and $C_1=0$  in \eqref{KdV_4} we also have equation
\begin{equation}\begin{gathered}
k^3\,V_{zz}+3\,a_0\,k\,V^2-\omega\,V=0.
\label{KdV_7}
\end{gathered}\end{equation}

Comparing two equations \eqref{KdV_6} and \eqref{KdV_7} we find that these equations are the same in case
\begin{equation}\begin{gathered}
a_0=2\,k^2,\,\quad \omega=k^3.
\label{KdV_8}
\end{gathered}\end{equation}

We obtain that the solution of equation \eqref{KdV_1} in the form
\begin{equation}\begin{gathered}
u(x,t)=2\,k^2\,V(z)=2\,k^2\,Q_z=2\,k^2\,(Q-Q^2).
\label{KdV_9}
\end{gathered}\end{equation}

This solution is the soliton of the Korteweg-de Vries equation in the form
\begin{equation}\begin{gathered}
u(x,t)=\frac{2\,k^2}{1+\exp{(k\,x-k\,x_0-k^3\,t)}}-
\frac{2\,k^2}{\left(1+\exp{(k\,x-k\,x_0-k^3\,t)}\right)^2},
\label{KdV_10}
\end{gathered}\end{equation}
where $x_0$ is an arbitrary constant. We have found the famous soliton solution of the Korteweg -- de Vries equation expressed via the logistic function too.

\section{Conclusion}

In this paper we have demonstrated that the logistic function is a solution of many nonlinear differential equations. We have illustrated that solutions of the the Burgers equation, the Burgers -- Huxley equation, the Burgers -- Fisher equation, the modified Korteweg -- de Vries equation with dissipation,  the Gardner equation, the Korteweg -- de Vries equation with source, the modified Korteweg -- de Vries equation with source and the generalized Kuramoto -- Sivashinsky equation are expressed via the logistic function. It is clear that the logistic function can be used for construction of exact solutions of many nonlinear differential equations.

We also presented the method for finding exact solutions of nonlinear partial differential equations. The basic idea of this method is the comparison of nonlinear ordinary differential equation obtained from PDE with the standard equation from the list of equations. Method is simple and do not require application of the symbolic calculations on computer.

\section{Acknowledgment}

This research was partially supported by Federal Target Programm
Research and Scientific-Pedagogical Personnel of Innovation
in Russian Federation on 2009-2013, by RFBR grant 11--01--00798--a
and "Researches and developments in priority directions of development of a scientifically-technological complex of Russia on 2007-2013".


\begin{thebibliography}{}



\bibitem{Gershenfield}
N.A. Gershenfield, The Nature of mathematical modeling, Cambridge, UK, Cambridge University Press, 1999.

\bibitem{Richards}
F.J. Richards, A flexible growth function for empirical use, J/ Exp. Bot., 10, 1959. 290--300.

\bibitem{KudrAMC13}
N.A. Kudryashov, Polynomials in logistic function and solitary waves of nonlinear differential equations, Appl. Math. Comput. 219 (2013)9245 -- 9253

\bibitem{Kudr12a}
N.A. Kudryashov, One method for finding exact solutions of nonlinear differential equations, Commun. Nonlinear Sci. Numer. Simulat. 17 (2012) 2248--2253.

\bibitem{KudrAMC13B}
N.A. Kudryashov, Quasi-exact solutions of the dissipative Kuramoto-Sivashinsky equation, Appl. Math. Comput. 219 (2013)9213 -- 9218

\bibitem{Kudr88}
N.A. Kudryashov, Exact soliton solutions of the generalized evolution equation of wave dynamics, PMM-J. Appl. Math. Mech. 52 (1988) 361--365.


\bibitem{Kudr90}
N.A. Kudryashov, Exact solutions of the generalized Kuramoto--Sivashinsky equation, Phys. Lett. A., 147 (1990) 287--291.

\bibitem{Kudr05a}  N.A. Kudryashov, Simplest equation method to look for exact solutions of nonlinear differential equations, Chaos Soliton Frac.  24 (2005) 1217--1231.


\bibitem{Kudr08a}N.A. Kudryashov,  N.B. Loguinova, Extended simplest equation method for nonlinear differential equations, Appl Math Comp  205 (2008) 396--402.


\bibitem{Kudr10a} N.A. Kudryashov, Meromorphic solutions of nonlinear ordinary differential equations, Commun Nonlinear Sci Numer Simulat. 15 (2010) 2778-2790.

\bibitem{Kabir01}M.M. Kabir, A.E.Y., KhajehAghdam, A.Y. Koma, Modified Kudryashov method for finding exact solitary wave solutions of higher order nonlinear equations, Mathematical methods in the Applied Sciences, 34(2)(2011) 213 - 219

\bibitem{Kabir02} M.M. Kabir, Modified Kudryashov method for Generalized forms of the nonlinear heat conduction equation, Internal Journal of Physical Sciences 6(25)(2011)6061-6064


\bibitem{Biswas}
A.J.M. Jaward, M.D. Petkovic, A. Biswas, Modified simple equation method for nonlinear evolution equations, Appl. Math. Comput. 217 (2010) 869--877.

\bibitem{Vitanov01}
N.K. Vitanov, Application of simplest equations of Bernoulli and Riccati kind for obtaining exact traveling-wave solutions for a class of PDEs with polynomial nonlinearity, Commun. Nonlinear Sci. Numer. Simulat. 15 (2010) 2050--2060.

\bibitem{Vitanov02}
N.K. Vitanov, Modified method of simplest equation: Powerful tool for obtaining exact and approximate traveling-wave solutions of nonlinear PDEs, Commun. Nonlinear Sci. Numer. Simulat. 16 (2011) 1176--1185.

\bibitem{Vitanov03}
N.K. Vitanov, On modified method of simplest equation for obtaining exact and approximate of nonlinear PDEs: The role of the simplest equation, Commun. Nonlinear Sci. Numer. Simulat. 16 (2011) 4215--4231.



\bibitem{Malfliet}
W. Malfliet, Solitary wave solutions of nonlinear wave equations, Am. J. Phys. 60 (1992) 650--654.

\bibitem{Hereman01}
W. Malfliet, W. Hereman, The tanh method: I. Exact solutions of nonlinear evolution and wave equations, Phys. Scripta. 54 (1996) 563--568.

\bibitem{Parkes01}
E.J. Parkes, B.R. Duffy, An automated tanh-function method for finding solitary wave solutions to nonlinear evolution equations, Comput. Phys. Commun. 98 (1996) 288--300.

\bibitem{Hereman02}
W. Malfliet, W. Hereman, The tanh method: II. Perturbation technique for conservative systems, Phys. Scripta. 54 (1996) 569--575.

\bibitem{Hereman03}
D. Baldwin, U. G\"oktas, W. Hereman, L. Hong, R.S. Martino, J.G. Miller, Symbolic computation of exact solutions expressible in hyperbolic and elliptic functions  for nonlinear PDEs, J. Symb. Comput. 37 (2004) 669--705.

\bibitem{Biswas01}A. Biswas, Solitary wave solution for the generalized Kawahara equation, Appl. Math. Lett. 22 (2009) 208--210.


\bibitem{Wang}
M.L. Wang, X. Li, J. Zhang, The G'/G--expansion method and evolution equation in mathematical physics, Phys. Lett. A. 372 (2008) 417--421.


\bibitem{Kudr09k}
N.A. Kudryashov, Seven common errors in finding exact solutions of nonlinear differential equations, Commun. Nonlinear Sci. Numer. Simulat. 14 (2009) 3503--3529.

\bibitem{Kudr10k}
N.A. Kudryashov, A note on the G'/G--expansion method, Appl. Math. Comput. 217 (2010) 1755--1758.




\bibitem{Kudr10n}
M.V. Demina, N.A. Kudryashov, From Laurent series to exact meromorphic solutions: The Kawahara equation, Phys. Lett. A. 374 (2010) 4023-–4029.

\bibitem{Kudr10ss}
M.V. Demina, N.A. Kudryashov, Explicit expressions for meromorphic solutions of autonomous nonlinear ordinary differential equations, Commun. Nonlinear Sci. Numer. Simulat. 16 (2011) 1127--1134.


\bibitem{Kudr11b}M.V. Demina, N.A. Kudryashov, On elliptic solutions of nonlinear ordinary differential equations, Applied Mathematics and Computation. 217 (2011) 9849--9853












\end{thebibliography}
\end{document}